\documentclass{aa}  

\usepackage{graphicx}
\usepackage{xcolor}
\usepackage{booktabs}
\usepackage{supertabular}
\usepackage[version=4]{mhchem}
\usepackage{ulem} 
\usepackage{longtable}
\usepackage{caption}
\usepackage{sidecap}
\usepackage{float}
\usepackage{placeins}
\usepackage{rotating}

\usepackage{txfonts}
\usepackage{hyperref}
\hypersetup{
    colorlinks=true,
    linkcolor= blue,
    filecolor=magenta,      
    urlcolor=blue,
    citecolor= blue,
}%
\begin{document}

   \title{ALMA view of the L1448-mm protostellar system on disk scales: CH$_3$OH and H$^{13}$CN as new disk wind tracers}


   \author{P. Nazari
          \inst{1}
          \and
          B. Tabone\inst{2}
          \and
          A. Ahmadi\inst{1}
          \and
          S. Cabrit\inst{3,4}
          \and
          E. F. van Dishoeck\inst{1,5}
          \and
          C. Codella\inst{6,4}
          \and
          J. Ferreira\inst{4}
          \and
          L. Podio\inst{6}
          \and
          {\L}. Tychoniec\inst{7}
          \and
          M. L. van Gelder\inst{1}
          }

   \institute{Leiden Observatory, Leiden University, P.O. Box 9513, 2300 RA Leiden, the Netherlands\\ 
        \email{nazari@strw.leidenuniv.nl}
         \and
         Universit\'{e} Paris-Saclay, CNRS, Institut d'Astrophysique Spatiale, 91405 Orsay, France
         \and
         Observatoire de Paris, PSL University, Sorbonne Universit\'{e}, CNRS UMR 8112, LERMA, 61 Avenue de l’Observatoire, 75014 Paris, France
         \and
         Univ. Grenoble Alpes, CNRS, IPAG, 38000 Grenoble, France
         \and
         Max Planck Institut f\"{u}r Extraterrestrische Physik (MPE), Giessenbachstrasse 1, 85748 Garching, Germany
         \and
         INAF, Osservatorio Astrofisico di Arcetri, Largo E. Fermi 5, I-50125 Firenze, Italy
         \and
         European Southern Observatory, Karl-Schwarzschild-Strasse 2, 85748 Garching, Germany
             }

   \date{Received 20 November 2023 / Accepted 19 February 2024}

 
  \abstract
  {Protostellar disks are known to accrete, however, the exact mechanism that extracts the angular momentum and drives accretion in the low-ionization "dead" region of the disk  is under debate. In recent years, magneto-hydrodynamic (MHD) disk winds have become a popular solution. Yet, observations of these winds require both high spatial resolution (${\sim}10$s\,au) and high sensitivity, which has resulted in only a handful of MHD disk wind candidates so far. In this work we present high angular resolution (${\sim}30$\,au) ALMA observations of the emblematic L1448-mm protostellar system and find suggestive evidence for an MHD disk wind. The disk seen in dust continuum (${\sim}0.9$\,mm) has a radius of ${\sim}23$\,au. Rotating infall signatures in H$^{13}$CO$^+$ indicate a central mass of $0.4\pm 0.1$\,M$_\odot$ and a centrifugal radius similar to the dust disk radius. Above the disk, we unveil rotation signatures in the outflow traced by H$^{13}$CN, CH$_3$OH, and SO lines and find a kinematical structure consistent with theoretical predictions for MHD disk winds. This is the first detection of an MHD disk wind candidate in H$^{13}$CN and CH$_3$OH. The wind launching region estimated from cold MHD wind theory extends out to the disk edge. The magnetic lever arm parameter would be  $\lambda_{\phi} \simeq 1.7$, in line with recent non-ideal MHD disk models. The estimated mass-loss rate is ${\sim}4$ times the protostellar accretion rate ($\dot{M}_{\rm acc} \simeq 2 \times 10^{-6} M_{\odot}/yr$) and suggests that the rotating wind could carry enough angular momentum to drive disk accretion.} 

   \keywords{Stars: protostars --
            ISM: molecules --
            Stars: winds, outflows --
            Techniques: interferometric
               }

   \maketitle
%

\section{Introduction} 
\label{sec:intro}

Star formation starts with a cloud of gas and dust which collapses as it rotates. Because of conservation of angular momentum, the envelope flattens and a disk forms. If angular momentum is not transported away, the disk cannot accrete and the star cannot grow (\citealt{Hartmann2016}). Therefore, angular momentum needs to be extracted from the disk by either turbulent stresses (\citealt{Shakura1973}; \citealt{Lynden1974}; \citealt{Balbus1998}) or magnetized disk winds (\citealt{Blandford1982}; \citealt{Ferreira1997}). Magneto-hydrodynamic (MHD) disk winds are expected to be a natural result of the vertical magnetic field in the disk inherited from the collapse (\citealt{Ferreira1997}; \citealt{Tomida2010}; \citealt{Bai2013}). These winds have become particularly popular in recent years as a viable solution to the angular momentum problem in disk regions $\simeq 1-50$ au, where ionization is too low to sustain MHD turbulence (see PPVII reviews by \citealt{Pascucci2023, Manara2023, Lesur2023}), and to explain disk demographics (\citealt{Tabone2022Short}). 


Yet, only a handful of spatially resolved observations of such disk wind candidates are available (\citealt{Launhardt2009}; \citealt{Bjerkeli2016}; \citealt{Hirota2017}; \citealt{Tabone2017}; \citealt{Lee2018}; \citealt{deValon2020, deValon2022}). The high sensitivity, angular and spectral resolution of the Atacama Large Millimeter/submillimeter Array (ALMA) are particularly needed to resolve the rotating signatures of disk winds and provide clues on their launch point \citep{Tabone2020}. Therefore, it is still an open question as to whether MHD disk winds are ubiquitous. It is not yet clear whether the low-velocity, wide-angle molecular outflow often surrounding the base of the high-velocity jet is simply a result of envelope entrainment by the central jet or if it mainly originates as an MHD wind launched from an extended region of the disk (see \citealt{deValon2022}; \citealt{Pascucci2023} for in-depth discussions).

\begin{table*}
\renewcommand{\arraystretch}{1.3}
    \caption{Targetted spectral lines and observational parameters}
    \label{tab:obs}
    \resizebox{\textwidth}{!}{\begin{tabular}{@{\extracolsep{1mm}}*{7}{l}}
          \toprule
          \toprule      
        Species & Frequency & $A_{\rm ij}$ & $E_{\rm up}$ & rms &  Spectral resolution & Beam\\ 
        &(GHz)&(s$^{-1}$)&(K)&(mJy\,beam$^{-1}$)&(km s$^{-1}$)&($\arcsec \times \arcsec$) \\
        \midrule     

CH$_3$OH ($2_{2,1}-3_{1,2}$) & 335.13357 & $2.7 \times 10^{-5}$ & 45 & 1.4 & 0.44 & $0.09\times 0.12$ \\
H$^{13}$CN (4-3) &  345.33977 & $1.9 \times 10^{-3}$ & 41 & 2.1 & 0.21 & $0.09\times 0.12$ \\
SO ($8_9-7_8$)&  346.52848 & $5.4 \times 10^{-4}$ & 79 & 2.2 & 0.21 & $0.09\times 0.12$ \\
SiO ($8_0-7_0$) &  347.33058 & $2.2 \times 10^{-3}$ & 75& 2.2 & 0.42 & $0.09\times 0.12$ \\
H$^{13}$CO$^+$ (4-3)&  346.99834 &  $3.3 \times 10^{-3}$ & 42  & 2.1 & 0.21 & $0.13\times 0.18$ \\
0.88\,mm continuum & 340.8399 & -- & -- & 0.25 & -- & $0.08 \times 0.12$\\

\bottomrule
        \end{tabular}}
        \tablefoot{The line rms is calculated in channels with widths given in the sixth column. The spectroscopic information of all the lines are taken from the Cologne Database for Molecular Spectroscopy (\citealt{Muller2001}; \citealt{Muller2005}).}
\end{table*}

In this paper we present high angular resolution ($\sim 0.1\arcsec = 30$\,au) and high sensitivity ALMA observations to characterize the disk of the emblematic L1448-mm protostellar system, analyze the base of its outflow, and present first evidence for its MHD disk wind. L1448-mm (also known as L1448-C and Per-emb 26) is a well-known Class 0 source located in the Perseus star-forming region ($d \simeq {\sim}300$\,pc; \citealt{Ortiz2018}), first betrayed by its spectacular outflow and jet (\citealt{Bachiller1990, Bachiller1991}), and with a luminosity of ${\sim}9$\,L$_{\odot}$ (\citealt{Andre2010}; \citealt{vantHoff2022}). Over the past 35 years, its dust continuum has been extensively studied at millimeter and centimeter wavelengths, notably by the CALYPSO program at modest angular resolution (${\sim}0.3-1\arcsec$; \citealt{Anderl2016}; \citealt{Maury2019}) and the VANDAM program at high angular resolution (${\sim}0.07-0.2\arcsec$; \citealt{Tobin2016}). The dust disk size inferred from uv visibility fit is ${\sim}0.16\arcsec$  (\citealt{Maury2019}). Complex organic molecules (COMs) are detected by CALYPSO over a larger region ${\sim}0.5\arcsec$ = 150 au, roughly consistent with a hot corino \citep{Belloche2020}. The angular momentum profile of the surrounding envelope shows a factor 10 decrease from 4000 to 1000~au suggestive of magnetic braking \citep{Gaudel2020}, making it a good candidate to search for an MHD disk wind. Its jet and outflow have also been studied intensively in SiO, CO, and SO (\citealt{Bachiller1995}; \citealt{Hirano2010}; \citealt{Podio2021}; \citealt{Tychoniec2021}; \citealt{Toledano2023}). However, these previous molecular line observations only had a modest angular resolution (${\sim}0.3\arcsec-1\arcsec$). Therefore, the new ALMA observations presented here are the first able to resolve the base of the outflow in this object and to constrain its potential MHD disk wind. 

\section{Observations}

L1448-mm was observed in Band 7 with the ALMA 12-m array during Cycle 8 combining a compact (C-3) and extended configuration (C-6) (PI: B. Tabone; project ID: 2021.1.01578.S). The key targeted species are SO, SiO, H$^{13}$CN, H$^{13}$CO$^+$, and CH$_3$OH (see summary in Table \ref{tab:obs}). The data were pipeline calibrated using CASA versions 6.2.1.7 and 6.4.1.12 (\citealt{McMullin2007}) and include nine spectral windows with frequencies ranging from ${\sim}333.786$\,GHz to ${\sim}347.557$\,GHz. Next we concatenated the measurement sets from the two C-3 and C-6 configurations. We then used CASA version 6.4.1.12 to perform continuum subtraction and imaging. The continuum visibilities were computed using carefully selected line-free channels in all spectral windows, and used to perform continuum subtraction in the spectral line uv-data. Next we used the \texttt{tclean} task for imaging. Given that the data are spatially resolved we set the \texttt{deconvolver} parameter to \texttt{multiscale}. For continuum and less extended species such as CH$_3$OH and H$^{13}$CN, we used a circular mask with a radius of $2\arcsec$ centered on the continuum peak ($\alpha_{2000}$ = $03^{\rm h}25^{\rm m}38.879^{\rm s}$, $\delta_{2000}$ =$+30^{\circ}44\arcmin05.210\arcsec$), encompassing all of the emission. For lines showing emission on larger scales, namely H$^{13}$CO$^+$, SiO, and SO, we manually masked the regions of interest. For all species except H$^{13}$CO$^+$ the \texttt{briggs} weighting was set to 0.5 to maximize angular resolution, yielding a clean beam of ${\sim}0.12\arcsec \times 0.09\arcsec$ at PA of ${\sim2^\circ-3^\circ}$, corresponding to a linear resolution of ${\sim}30$\,au at the distance of Perseus. Because H$^{13}$CO$^+$ line is relatively weak, we used a \texttt{briggs} weighting of 2.0 to increase its signal to noise. The resulting clean beam FWHM is ${\sim}0.18\arcsec \times 0.13\arcsec$ at a PA of ${\sim}-4^\circ$, corresponding to ${\sim}45$\,au in Perseus. The maximum recoverable scale for the observations is ${\sim}6\arcsec$ (i.e., ${\sim} 1800$\,au).

The observational properties for the final cube of each molecule are summarized in Table \ref{tab:obs}. 
The rms varies from ${\sim}1.5$\,mJy\,beam$^{-1}$ to ${\sim}2$\,mJy\,beam$^{-1}$ in channel widths of 0.2 or 0.4\,km\,s$^{-1}$ (see Table \ref{tab:obs}). In the following analysis, we adopt a systemic velocity of V$_{\rm lsr}$ = 5.3\,km\,s$^{-1}$, inferred from the emission lines of CH$_3$OH.





\section{Anatomy of the L1448-mm system}

\begin{SCfigure*}
    \includegraphics[width=11cm]{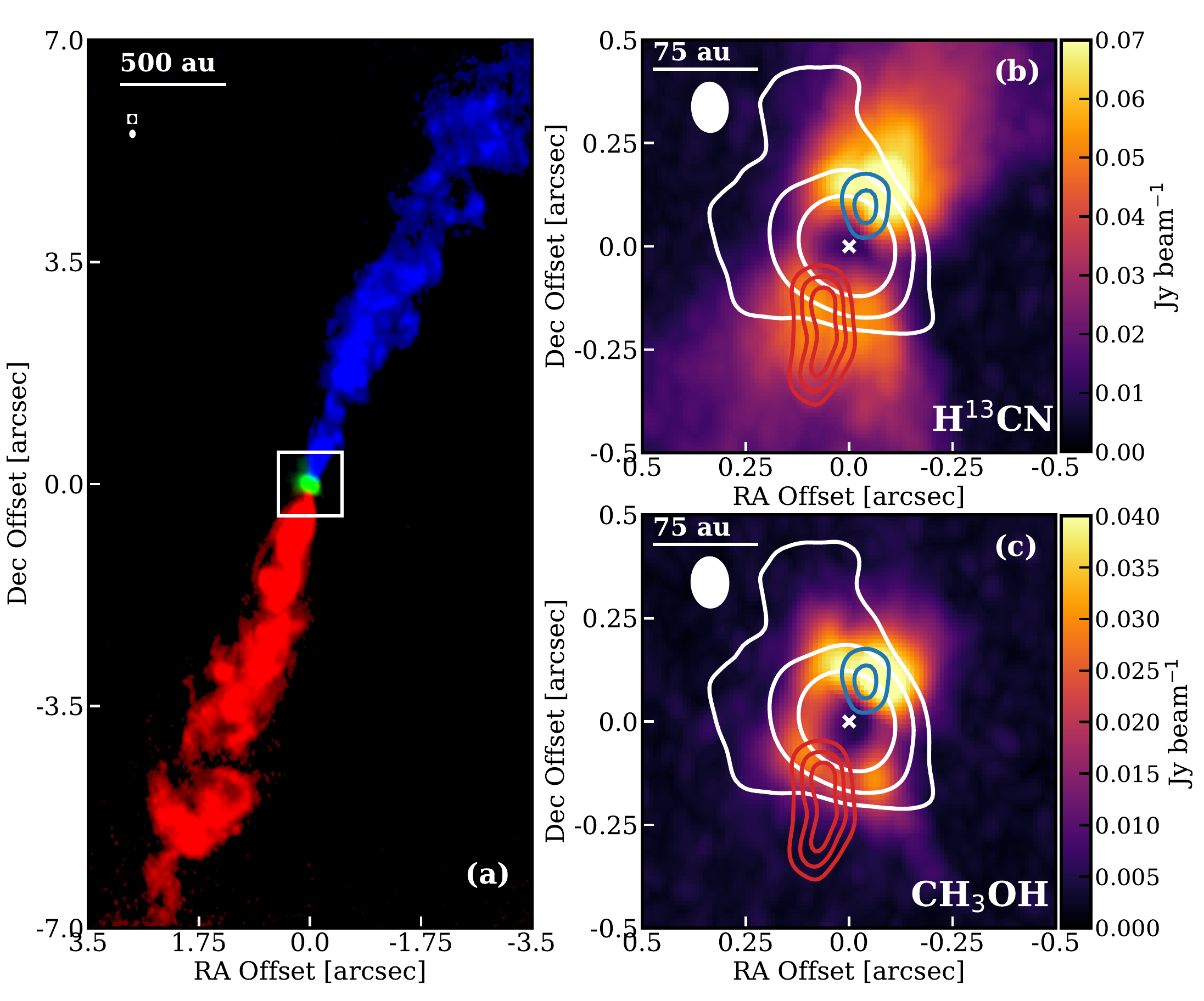}
    \caption{Overview of the L1448-mm system as seen by ALMA. Panel (a) shows an overview of the region. The red and blue lobes of the jet are indicated by SiO (8-7) peak intensity map. Green corresponds to the ${\sim}0.9$\,mm continuum. The beams of SiO (empty) and continuum (filled) are shown on the top left. Panel (b) shows a zoom-in of H$^{13}$CN (4-3) peak intensity map with continuum contours in white at [10, 30, 100]$\sigma_{\rm cont}$ with $\sigma_{\rm cont} = 0.25$\,mJy\,beam$^{-1}$. Panel (c) shows the same for CH$_3$OH (2$_{2,1,0}-3_{1,2,0}$). In panels (b) and (c) the peak of the continuum is indicated by a cross and the integrated intensity maps of high-velocity bullets of SiO (65-75\,km\,s$^{-1}$) are shown in red and blue contours (at [10, 15, 20]$\sigma_{\rm SiO, mom0}$ with $\sigma_{\rm SiO, mom0}=6$\,mJy\,beam$^{-1}$\,km\,s$^{-1}$).} 
    \label{fig:region}
\end{SCfigure*}

Figure \ref{fig:region}a presents an overview of the L1448-mm system, dominated at large scale by a spectacular jet seen in SiO emission. Our high angular resolution observations allow us to peer into the structure of the system on disk scales for the first time, as shown in Fig. \ref{fig:region}b and c. The innermost SiO knots at $\pm$ 0.15\arcsec from the source define a current jet PA = -18\degr, that we will adopt in this work.

\subsection{Dust disk size and mass}
\label{sec:disk}

\begin{figure}
  \resizebox{\columnwidth}{!}{\includegraphics{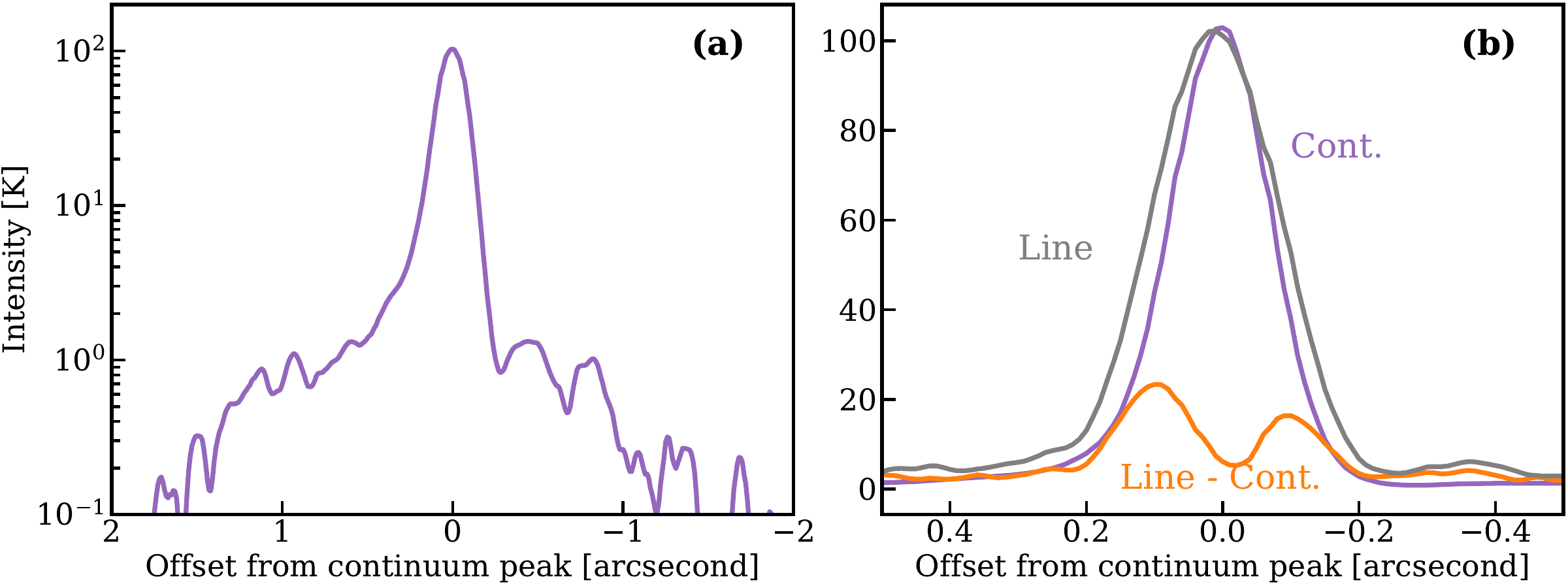}}
  \caption{Intensity cuts perpendicular to the jet axis. (a) Continuum intensity profile on larger scale, showing the bright central disk on top of an extended envelope pedestal. (b) Intensity profiles on disk-scales of continuum (purple), CH$_3$OH before continuum subtraction (gray), and CH$_3$OH after continuum subtraction (orange), showing the artificial formation of a central "hole" surrounded by a faint ring.}
  \label{fig:intensity}
\end{figure}

The continuum intensity contours show an extended plateau, tracing the envelope, and a sudden increase of ${\sim}2$ orders of magnitude in the inner ${\sim}0.2\arcsec$, best seen in an intensity cut across the continuum in Fig. \ref{fig:intensity}a, which is attributed to a disk. Assuming a Gaussian model for the disk, we fitted for the disk size and its flux in the uv visibility plane. We find a disk FWHM of $0.15\arcsec\pm 0.01\arcsec \times 0.12\arcsec\pm 0.01\arcsec$, in line with previous disk size estimates (\citealt{Maury2019}; \citealt{Toledano2023}). The integrated flux from the fit is ${\sim} 266 \pm 18$\,mJy with the disk PA of ${\sim}65^{\circ}\pm3^{\circ}$. The disk PA found from this method agrees with being roughly perpendicular to the jet. We also found that fitting 1D Gaussians to intensity cuts across the continuum in perpendicular and parallel directions to the jet axis, and deconvolving the fitted FWHMs from the beam (see Appendix D of \citealt{Podio2021}), provides the disk FWHMs fully consistent with the measured disk size from the uv visibility fit. Using our estimated disk major and minor axes (${\sim}45 \pm 3 \times 36 \pm 3$\,au), a lower limit on inclination angle from the line of sight of $i {\sim}30^{\circ}-35^{\circ}$ is found assuming a geometrically thin disk. This is significantly smaller than the previously adopted inclination of 70\degr \citep{Girart2001,Gaudel2020}, but is in agreement with more recent high-resolution proper motions of SiO jet knots by \cite{Yoshida2021}; they find $i \sim 30^{\circ}$ for the innermost, brightest knots RI-a and RI-b (see their Fig.~6) and $i \sim 44^{\circ}$ on average in the red lobe (of much higher signal to noise than the blue one). Therefore we will adopt $i = 30-44^{\circ}$ in the following.


The maximum brightness temperature of the continuum is high, $\simeq 110~K$, and exceeds the sublimation temperatures of methanol ices (${\sim}70-100$\,K; \citealt{Ferrero2020}; \citealt{Ceccarelli2022}; \citealt{Minissale2022}). Yet, peak intensity maps of CH$_3$OH (2$_{2,1,0}-3_{1,2,0}$) line shows a hole centered on the continuum peak on disk scales (see Figs. \ref{fig:region}b,c). Particularly, the intensity cuts of methanol in Fig. \ref{fig:intensity}b show that before continuum subtraction, the line intensity is centrally peaked, while after continuum subtraction the line intensity drops to near zero in the center. Given that the peak brightness temperature of the dust emission (${\sim}100$\,K; Fig. \ref{fig:intensity}) is consistent with typical dust temperatures in the inner ${\lesssim} 30$\,au of protostellar systems from radiative transfer models (e.g., \citealt{Nazari2022Lowmass}), it is reasonable to assume that the dust is optically thick on source in the disk, as also suggested by the spectral index of 1.8 observed for the compact disk component between 0.88\,mm (integrated flux of ${\sim}266$\,mJy, see above) and 1.3\,mm ($130\pm 5$ mJy), and consistent with the slope of $2\pm0.2$ previously reported between 1.3\,mm and 2.7\,mm at lower angular resolution (\citealt{Maury2019}; \citealt{Toledano2023}).


Therefore, the hole is likely due to dust optical depth effects. However, it cannot be dust attenuation in the envelope, as in that case, the hole would be on larger scales (${\ge}0.3\arcsec$, see Fig. \ref{fig:region}c). Given that the hole and the dust disk have similar sizes, the effect arises because of dust extinction within the disk (\citealt{DeSimone2020}; \citealt{vanGelder2022}) and/or continuum over-subtraction (\citealt{Boehler2017}; \citealt{Weaver2018}; \citealt{Rosotti2021}). However, it cannot be due only to dust extinction in the disk because, for the moderate disk inclination of L1448-mm (30\degr-50\degr) some methanol emission on top of the disk (i.e., between the disk and the observer) should still be present. 


Therefore, continuum over-subtraction is likely causing this hole in methanol emission. This effect naturally arises when the continuum is optically thick and the gas located on top of the dust has a similar excitation temperature as the dust, $T_{\rm ex} \simeq T_{\rm d}$. Denoting the line opacity as $\tau_{\nu}$, the emerging intensity is given by:
\begin{equation}
    I_{\nu} = B_{\nu}(T_{\rm d}) e^{-\tau_{\nu}} + B_{\nu}(T_{\rm ex}) (1-e^{-\tau_{\nu}}),
\end{equation}
where the first term accounts for the underlying disk dust emission attenuated by the gas on top of it and the second term accounts for the intrinsic gas line emission. Subtracting the continuum, which is $I^{\rm{cont}}_{\nu} = B_{\nu}(T_{\rm d})$, we obtain
\begin{equation}
    I_{\nu} - I^{\rm cont}_{\nu} = \left[B_{\nu}(T_{\rm ex})- B_{\nu}(T_{\rm d})\right] (1-e^{-\tau_{\nu}}),
\end{equation}
\noindent which vanishes when $T_{\rm ex} \simeq T_{\rm d}$, regardless of the line opacity $\tau_{\nu}$. The amount of gas can be arbitrarily large, its emission will always cancel out after continuum subtraction, leaving a hole with no detectable line signal there. A more intuitive way to understand this effect is to note that the gas absorbs as many photons as it emits. Therefore no spectral line can appear on top of the continuum.


This is further supported by the radiative transfer models of \cite{Nazari2022Lowmass} which showed that a considerable decrease in methanol emission is only possible if a disk with optically thick dust is present and through the continuum over-subtraction effect. Therefore, we conclude that gas-phase methanol is likely present in the disk inside the hole but cannot be observed at millimeter wavelengths. 
We however note that extinction can still play a role on larger scales and the dust in the envelope could be decreasing the emission from the redshifted outflow lobe, because it is fainter than the blueshifted one which is piercing through the envelope and coming toward us. Similar effects are likely responsible for the similar emission morphology of H$^{13}$CN.

The disk mass is quite uncertain, but is probably below 0.1\,M$_\sun$ and possibly as small as 0.01\,M$_\sun$. The 9\,mm fluxes, corrected for free-free contribution, lie below the extrapolated 1.3\,mm flux assuming a slope of 2; therefore, the 9\,mm emission seems partly optically thin, and can be used to estimate the disk mass, $M_{\rm d}$, with the well-known formula (\citealt{Hildebrand1983}; \citealt{Tychoniec2018})
\begin{equation}
    M_{\rm d} = \frac{D^2 F_\nu}{B_\nu(T_{\rm d}) \kappa_\nu},
\end{equation}
where $D$ is the source distance, $F_\nu$ the integrated continuum flux, $T_d$ the average dust disk temperature, and $\kappa_\nu$ is the dust opacity (including gas) at the considered frequency. \citealt{Tychoniec2018} obtained a disk mass of 0.27$M_\odot$, using a typical disk dust temperature of 30~K and dust opacity law (extrapolated from 1.3\,mm to 9\,mm with $\kappa_\nu \propto \nu$) of \cite{Ossenkopf1994} for grains with thin ice mantles coagulated at densities $10^6$ cm$^{-3}$. The higher dust disk temperature of 100~K revealed by our observations reduces the mass to 0.08 $M_\odot$. Moreover, in such a warm and compact disk, it seems more appropriate to consider dust aggregates without ice mantles, and coagulation at higher gas densities $\ge 10^8$ cm$^{-3}$. Using the corresponding model in \cite{Ossenkopf1994} increases $\kappa_\nu$ by a factor 6.5, decreasing the disk mass further to 0.01$M_\odot$.

\subsection{Infall, stellar mass, centrifugal radius, and accretion rate}
\label{sec:envelope}

\begin{figure}
\centering
  \resizebox{0.9\columnwidth}{!}{\includegraphics{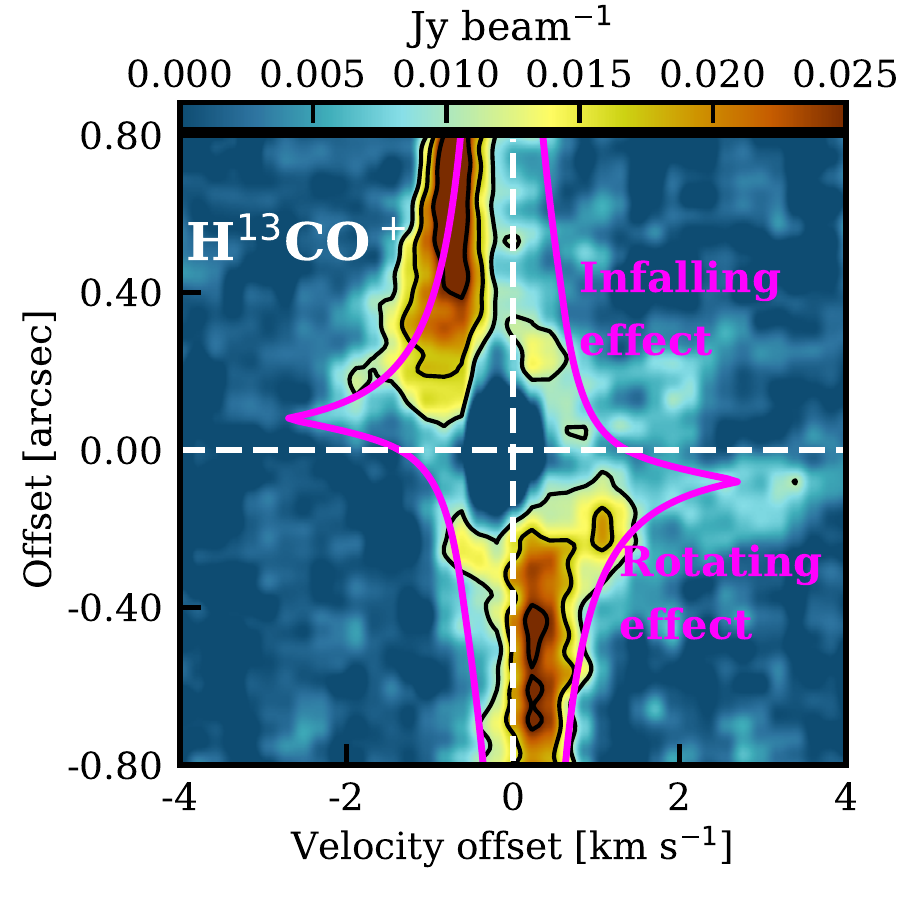}}
  \caption{P-V diagram of H$^{13}$CO$^+$ (4-3) perpendicular to the jet axis, averaged over a slit width of ${\sim}1$ beam. Black contours show the [5, 8, 12]$\sigma$ levels, where $\sigma$ is given in Table \ref{tab:obs}. The magenta curves indicate an infalling rotating envelope model with $l_{\rm env}$ of 130\,au\,km\,s$^{-1}$, $M_\star = 0.4$\,M$_{\odot}$, inclination angle of $i = 30^\circ$ to the line of sight, and centrifugal barrier $r_c = 24$ au.}
  \label{fig:PVs_three}
\end{figure}

The larger scale envelope is traced by H$^{13}$CO$^+$ for this source (\citealt{vantHoff2022}). Figure \ref{fig:HCO} presents the channel maps of H$^{13}$CO$^+$. Already in Fig. \ref{fig:HCO} some signs of rotation in the envelope are present, as well as infall (superposed blue and red emission at the same position). To investigate this further, Fig. \ref{fig:PVs_three} presents its position-velocity (P-V) diagram over a cut perpendicular to the jet axis through the continuum peak. Figure \ref{fig:PVs_three} shows that the kinematics of H$^{13}$CO$^+$ along the disk major axis on scales 0.1--0.8$\arcsec = $30--240\,au is well reproduced by a ballistic model of rotating and infalling flattened envelope (see \citealt{Sakai2014}; \citealt{Lee2017Infall} for model details) with a central mass of ${\sim}0.4\pm0.1$\,M$_{\odot}$ (constrained by the infall pattern of blue plus red superposition at low $V$ and large radii) and a constant de-projected specific angular momentum of $l_{\rm env} \simeq 130\pm20$\,au\,km\,s$^{-1}$ (constrained by the pure rotating pattern at radial velocities $V > 2$ km\,s$^{-1}$), considering inclination angles ${\sim}30^\circ-44^\circ$ (see Sect. 3.1). In the rest of this work, we consider two protostellar masses ($M_\star$) of 0.3\,M$_{\odot}$ and 0.5\,M$_{\odot}$ to cover the measured range of masses including its uncertainty\footnote{the effect of a disk mass of $0.1 M_\odot$ is discussed in Sect. 4.2}. \cite{Gaudel2020} found a similar value on the same scales, $l_{\rm env} \simeq 120\pm20$\,au\,km\,s$^{-1}$, despite their use of a much higher inclination $i =$ 70\degr. We checked that their measured $V(R)$ agree very well with the brightest "branches" of our PV cut in H$^{13}$CO$^+$. The fainter "branches" of opposite velocity sign (which signify infall) could not be retrieved with their method\footnote{They fitted a single position centroid in each velocity channel (at radii $<1 \arcsec$), or a single velocity centroid (moment~1) at each position (at radii $1-3.5\arcsec$).}. The neglect of infall motions thus increases the contribution of rotation to the $V(R)$ curve, leading to a larger projected angular momentum $l_{\rm env} \sin{i}$. The use of $i = 70^{\circ}$ by \citealt{Gaudel2020} to deproject, instead of $i = 30^{\circ}-44^{\circ}$ here, has the opposite effect, leading to a deprojected $l_{\rm env}$ similar to ours. 

The corresponding centrifugal barrier radius, $R_{\rm b} =$ $l_{\rm env}^2/(2GM_{\star}) \sim 24\pm 9$\,au, agrees with the dust disk radius (${\sim}23$\,au; see Sect. \ref{sec:disk}) as also found for L1527 and HH212 (\citealt{Sakai2014}; \citealt{Lee2017Infall}). The protostellar mass can be used to estimate the accretion rate onto the star, $\dot{M}_{\rm acc}$, assuming that the bolometric luminosity is dominated by the accretion luminosity, ${L}_{\rm acc} \simeq G M_{\star} \dot{M}_{\rm acc} / R_{\star}$ (\citealt{Hartmann1998}). For $L_{\rm acc}$ = 9\,L$_{\odot}$, $M_{\star}=0.3-0.5$\,M$_{\odot}$, and taking the appropriate stellar radius for an accreting protostar of this mass, $R_{\star}=3$\,R$_{\odot}$ (\citealt{Stahler1988}), we find $\dot{M}_{\rm acc} \simeq (2-3) \times 10^{-6}$\,M$_{\odot}$\,yr$^{-1}$. This amounts to 10 times the mass-flux of the high-velocity jet in L1448-mm, inferred from [OI]63\,$\mu$m maps (\citealt{Nisini2015}), a result fully consistent with the typical ejection/accretion ratio $\simeq$ 10\% observed in protostellar jets \citep[see e.g.][]{Lee2020review}. 

\subsection{Rotating small-scale outflow}
\label{sec:outflow}

\begin{figure*}
    \centering
    \includegraphics[width=17cm]{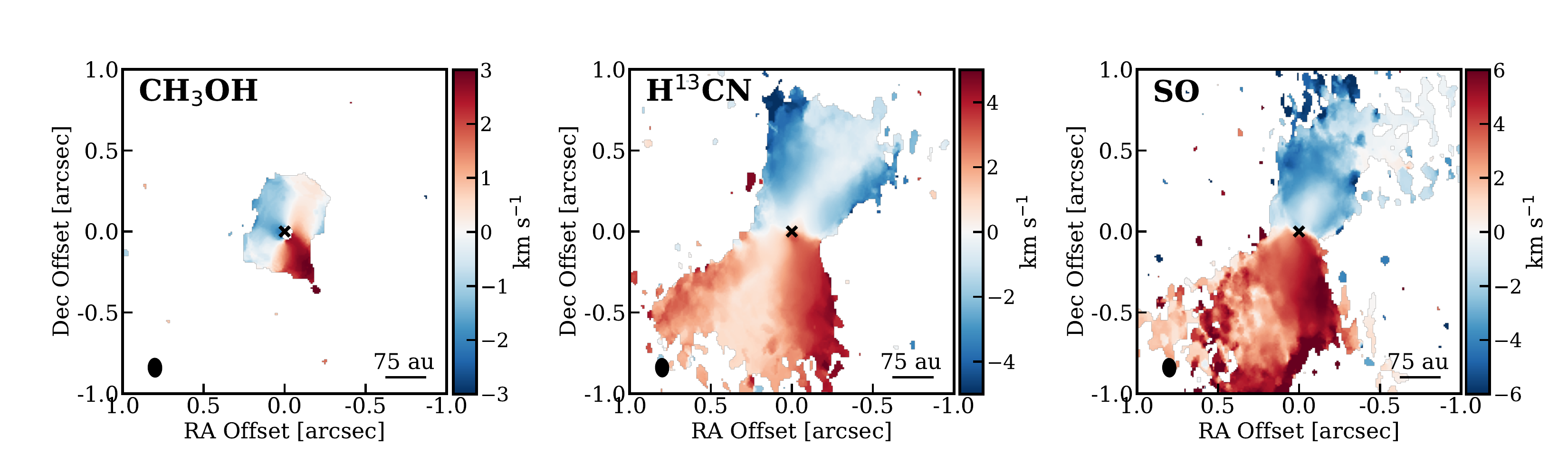}
    \caption{Moment 1 (intensity-weighted velocity) maps of CH$_3$OH (2$_{2,1,0}-3_{1,2,0}$), H$^{13}$CN (4-3), and SO (8$_9-7_8$) demonstrating rotation in their low-velocity outflow component. The velocity ranges used to compute the maps are $\pm4$, $\pm5$, and $\pm8$ km\,s$^{-1}$ for CH$_3$OH, H$^{13}$CN and SO, respectively and a cut at ${\sim}3\sigma$ is made. The black cross in all panels indicates the continuum peak.} 
    \label{fig:mom1}
\end{figure*}

Moment 1 maps and transverse P-V diagrams of the selected small-scale outflow tracers (CH$_3$OH, H$^{13}$CN, and SO) are presented in Figs. \ref{fig:mom1} and \ref{fig:PVs}, while their channel maps are presented in \ref{fig:chans_all}. Of all the three tracers, methanol exhibits the most compact emission. This is likely due to methanol mainly forming in prestellar ices and later sublimating close to the protostar at temperatures of ${\gtrsim}100$\,K (\citealt{Minissale2022}), while SO and HCN can efficiently form in the gas phase in disk winds \citep{Panoglou2012,Gressel2020}. 


The transverse centroid gradient in methanol Moment 1 map, and the tilt in its transverse PV-cuts, both show clear signatures of rotation in the same direction as the envelope. However, the emission is elongated along the jet axis, which is inconsistent with disk emission. Moreover, the Moment 1 map reveals a global velocity shift of $\pm$0.5~km~s$^{-1}$ between the red (SE) and blue (NW) lobes, consistent with a slow outflowing motion. These results demonstrate that methanol emission in L1448-mm traces the base of the outflow and not the inner envelope or the rotating disk. This supports the proposal of \cite{Codella2018} who suggested that complex organics in HH 212 could be a part of the outflow, and revises the classical picture of hot corinos as tracing the infalling envelope material \citep{Ceccarelli2004}.

\cite{Belloche2020} argued that the emission size of complex organic molecules in L1448-mm can be marginally explained by a hot corino scenario (i.e., thermal desorption close to the protostar). This would agree with the prediction of 2D radiative transfer models of \cite{Nazari2022Lowmass} for disk plus envelope structures including outflow cavities. They showed methanol emission along and outside the outflow cavity walls due to thermal desorption, with a similar size and flux as observed here. However, whether the methanol emission inside of the outflow itself is due to thermal desorption or mechanical ice sputtering due to ion-neutral drift is unclear.


H$^{13}$CN emission also traces the base of the outflow but extends further away from the disk, up to a projected distance of $z_{\rm proj}{\sim}150$~au. Its Moment 1 map, and the tilt at low velocity in transverse P-V cuts, again reveal clear rotation signatures consistent with the rotation direction of the envelope. In contrast with methanol, however, the PV-cuts in H$^{13}$CN at small vertical distances $z_{\rm proj} \simeq 30-60$~au also trace faster gas closer to the jet axis, suggestive of an "onion-like" nested velocity structure inside the low-velocity rotating flow. At larger distance ($z_{\rm proj}= 120$~au), this faster component disappears and the P-V cut exhibits an empty ring-like morphology, pointing toward a thinner shell of outflowing gas, where the rotation signature (tilt) clearly persists. Such a ring-like structure has been observed previously in other disk wind candidates such as HH30 and Orion Source I (\citealt{Louvet2018}; \citealt{Hirota2017}; \citealt{lopez_vazquez2020}). 

SO has previously been identified in HH 212 as a promising tracer of disk winds \citep{Tabone2017}. Here, comparison of the PV cuts in Fig. \ref{fig:PVs} confirms that SO is tracing the same rotating small-scale outflow as H$^{13}$CN. However, the emission appears less extended in the transverse direction and more complex compared with H$^{13}$CN (eg. the presence of both blue- and red-shifted emission in the NW PV-cuts at $z_{\rm proj} = 30-90$~au). This is likely due to a combination of chemical and optical depth effects, and to a larger contribution from jet bowshocks in SO. In the following, we therefore focus on CH$_3$OH and H$^{13}$CN.

\section{Evidence for an MHD disk wind}
\label{sec:wind}

\subsection{Wind morphology and kinematics}

\begin{figure*}
    \centering
    \includegraphics[width=19cm]{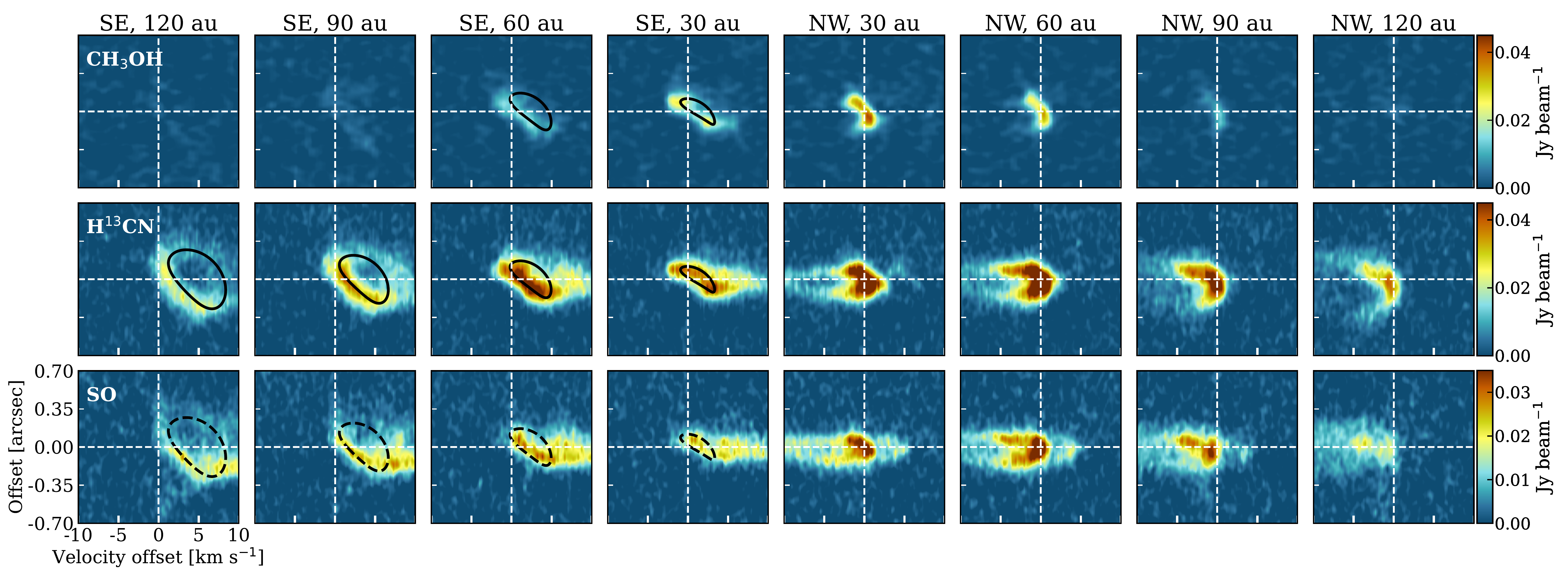}
    \caption{Transverse position-velocity (PV) diagrams of CH$_3$OH (2$_{2,1,0}-3_{1,2,0}$),  H$^{13}$CN (4-3), and SO (8$_9-7_8$) perpendicular to the jet axis, at projected offsets $z_{\rm proj} {\sim}30$\,au, $60$\,au, $90$\,au, and $120$\,au from the continuum peak, and averaged over a slit width of ${\sim}1$ beam. Cuts through the red (resp. blue) lobe are labelled as SE for South-East (resp. NW for North-West). Black solid curves show predictions for a thin shell model of rotating conical flow with a base radius of ${\sim}19$\,au, obtained by fitting the SE PV-cuts of CH$_3$OH and H$^{13}$CN (see green curves in Fig. \ref{fig:all_v_r_z}). The same model is overlaid in dashed lines on SE PV cuts of SO for comparison.} 
    \label{fig:PVs}
\end{figure*}

The rotation signatures in the outflow of L1448-mm, its small anchor radius, and its nested velocity structure (see Sect. \ref{sec:outflow}), are strongly suggestive of an MHD disk wind. 
In fact the shape of the P-V diagrams is well in line with the synthetic predictions of extended MHD disk winds, with faster gas launched closer to the protostar \citep{Tabone2020}. In order to more quantitatively test this claim, we analyze the P-V diagrams of H$^{13}$CN and CH$_3$OH focusing on the redshifted lobe. The PV-cuts there are more dominated by the low-velocity rotating component than in the blueshifted lobe, and show a clearer ring pattern at $z_{\rm proj}= 120$~au, allowing for a more robust fitting of the outermost emitting layer.

We use a method independent from simulations of MHD disk winds to retrieve the spatio-kinematical structure of the outer outflow from the observed P-V diagrams. Our model assumes a thin, axisymmetric rotating shell of gas to retrieve the flow shape $r(z)$, $v_{\phi}$ (rotational velocity), and $v_z$ (velocity along the axis). The transverse velocity $v_{r}$ can also be inferred assuming flow along the cone. Details of our method are explained in Appendix \ref{sec:toy_model} (see also, \citealp{Louvet2018,Tabone2020,deValon2022}).  




The black curves in Figure \ref{fig:PVs} show the location of emission expected from the thin shell model at different projected distances from the protostar ($z_{\rm proj}$). We find that the morphology of the brightest features in the observed P-V diagrams match well with the thin shell model. Overall, the shape of the outflow is well fitted by a conical flow with a half-opening angle of $i\simeq 13^{\circ}$, with its base at ${\sim}19$\,au from the central protostar. The observed P-V diagram of H$^{13}$CN is notably well described at $z_{\rm proj} = 120$\,au by a cone-like structure for which the base is traced by CH$_3$OH emission. 
At $z_{\rm proj} \lesssim 60$\,au, H$^{13}$CN P-V cuts show additional emission closer to the jet axis and at higher velocity that is not reproduced by the model.  At these locations, our thin shell model provides only the shape and kinematics of the outermost wind streamline emitting in this tracer.

We find that the retrieved axial velocity of the flow increases with distance, from $v_z  \simeq  1~$km~s$^{-1}$ at its base (traced by both methanol and H$^{13}$CN) to about 4~km~s$^{-1}$ at a deprojected distance of $z{\simeq}260$~au (traced only by H$^{13}$CN; Fig. \ref{fig:all_v_r_z}). The rotation velocity is found to be constant and about $v_{\phi} \simeq 4 \pm 2~$km~s$^{-1}$. Given the large uncertainties on $v_{\phi}$, we infer a tentative increase in the specific angular momentum with distance. 

\begin{figure}
  \resizebox{\columnwidth}{!}{\includegraphics{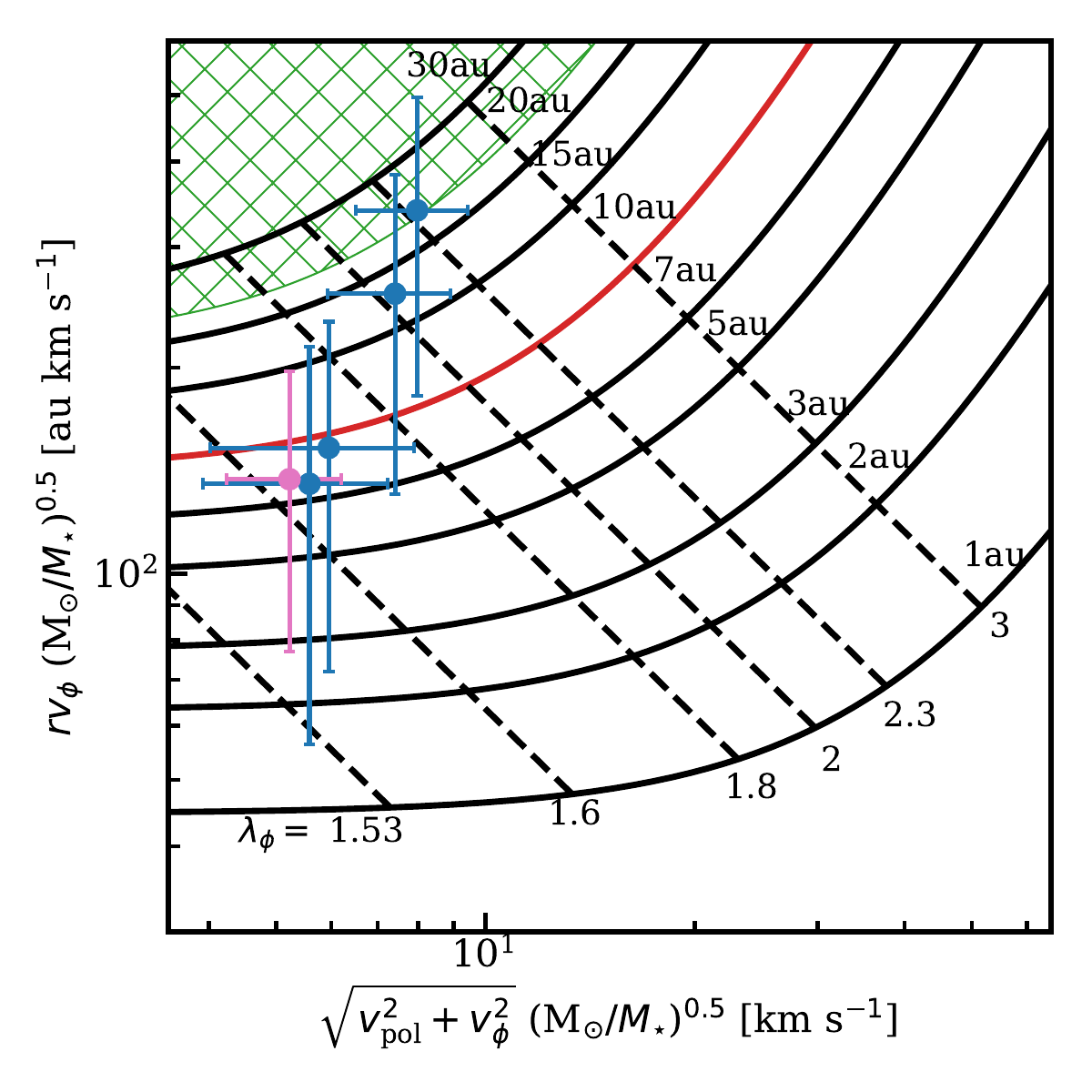}}
  \caption{Diagnostic diagram of MHD disk-wind launch point and magnetic lever arm, updated from \cite{Ferreira2006}. Specific angular momentum as a function of the total gas velocity. Data points at different altitudes from Fig. \ref{fig:all_v_r_z} are shown in blue (H$^{13}$CN) and pink (CH$_3$OH), where a protostellar mass of $0.5$\,M$_{\odot}$ was adopted. The black grid in the background shows the relation between these two quantities predicted 
  when gravitational potential is negligible, for different values of the wind launching radius $r_0$ (solid curves, Eq. \ref{eq:anderson_MHD}) and magnetic lever arm parameter $\lambda_\phi$ (dashed curves, Eq. \ref{eq:lambda_phi}). We note that in contrast to the first version of this diagram introduced by \cite{Ferreira2006}, the x-axis here plots the total gas velocity instead of $v_{\rm pol}$, since the measured $v_{\phi}$ is of the same order as $v_{\rm pol}$. The region with $r_0$ above the dust disk radius of 23\,au is hatched in green.}
  \label{fig:diagnostic}
\end{figure} 


\subsection{Wind launching point and magnetic lever arm parameter}

The retrieved kinematics of the wind allows us to infer the launching point of the wind $r_0$ using the "Anderson relation" (Eq. 4 of \citealt{Anderson2003}), given by
\begin{equation}
    2rv_{\phi}\Omega_0 = v_{\rm pol}^2 + v_{\phi}^2 + 3\Omega_0^2 r_0^2.
    \label{eq:anderson_MHD}
\end{equation}
\noindent This relation is valid at any point in a cold, steady, axisymmetric MHD disk wind where gravitational potential can be neglected, and directly follows from the conservation of energy and angular momentum \citep{Anderson2003}. In Eq. \eqref{eq:anderson_MHD}, $r$ is the radial distance from the jet axis, $v_{\phi}$ the rotational velocity, $v_{\rm pol}$ the poloidal velocity, and $\Omega_0$ is the orbital frequency at the launching radius $r_0$  ($\sim (GM_{\star}/r_0^3)^{1/2}$). 

Our results are summarized in Fig. \ref{fig:diagnostic} where we plot our CH$_3$OH and H$^{13}$CN measurements at various $z_{\rm proj}$ (from Fig. \ref{fig:all_v_r_z}) on top of the predicted relation between $rv_{\phi}$ and $v_{\rm tot} = (v_{\rm pol}^2 + v_{\phi}^2)^{1/2}$ from Eq. \eqref{eq:anderson_MHD}, for various values of $r_0$ (solid curves). We considered a range of stellar mass, $M_{\star}$, between 0.3-0.5\,M$_{\odot}$, as inferred from envelope infall kinematics (Sect. \ref{sec:envelope}). Figure \ref{fig:diagnostic} assumes a protostellar mass of 0.5\,M$_{\odot}$ and Fig. \ref{fig:diagnostic_03} shows the same diagram assuming a smaller $M_{\star} = 0.3\,M_{\odot}$. The conclusions are unchanged. Even if the disk mass was as large as 0.1\,M$_{\odot}$ (the upper limit estimated in Sect. 3.1), and the protostellar mass as small as 0.2\,M$_{\odot}$, the inferred launch radii would only increase by ${\sim}$1-2\,au.


Based on Figs. \ref{fig:diagnostic} and \ref{fig:diagnostic_03} we find that the launching radii inferred at the various distances $z_{\rm proj}$ roughly agree with each other at $r_0 \simeq 10^{+13}_{-8}~$au. Here, 10\,au is approximately the average of the two medians of $r_{0}$ for $M_{\star}$ of 0.3\,M$_{\odot}$ and 0.5\,M$_{\odot}$. The uncertainties on this value are based on the range that is seen in the data points (including their error bars) and up to the radii below the disk radius of 23\,au. We note that the inferred launching radius corresponds to the outermost emitting streamline of the wind, since we retrieved the wind kinematics from the outer boundary of the outflow in PV-cuts. This is consistent with the determined dust disk radius of ${\sim}23$\,au (Sect. \ref{sec:disk}) and the extrapolated anchor radius of the cone-like shape of the outflow from our best-fit model to PV diagrams of Fig. \ref{fig:PVs} (${\sim}19$\,au; see Appendix \ref{sec:toy_model}). It excludes an X-wind origin, with launching radius $\lesssim0.1$\,au. Including the gravitational potential term ($\Phi_g$ in Eq. 3 of \citealt{Anderson2003}) would only affect the launching radius by ${\sim}1-5$\,au, which is within the uncertainties of our $r_0$ estimates. 
We also find a tentative increase in the wind launching radius from $r_0 \simeq 10~$au to $r_0 \simeq 20~$au moving further away from the disk. 
One possibility is that the wind at large distances is slightly perturbed and compressed by jet bow-shocks \citep{Tabone2018}, 
highlighting outer streamlines that remained unseen at the base of the outflow due to insufficient excitation or chemical abundance of HCN. 
In fact, a perturbation by the jet would also explain why the outflow is resolved as a relatively thin shell at large distance from the disk ($z_p \gtrsim 90$), while it would be constantly refilled by the disk wind at low altitudes \citep{Tabone2018}.

A lower limit to the magnetic lever arm parameter can also be inferred as $\lambda_\phi \equiv r v_\phi / (\Omega_0 r_0^2)$ \citep{Ferreira2006}. Combining this expression with Eq. \eqref{eq:anderson_MHD} yields $v_{\rm tot}^2 = v_{\rm pol}^2 + v_{\phi}^2 = (2\lambda_\phi-3)\Omega_0^2 r_0^2$, hence
\begin{equation}
    (rv_{\phi} \times v_{\rm tot})/GM_\star = \lambda_\phi \sqrt{2\lambda_\phi -3}.
    \label{eq:lambda_phi}
\end{equation}
This predicted inverse relation between $rv_{\phi}$ and $v_{\rm tot}$ is plotted as dashed curves in Figs. \ref{fig:diagnostic} and \ref{fig:diagnostic_03}, for various values of $\lambda_\phi$. Comparison with the observed data points in L1448-mm indicates relatively low values of $\lambda_\phi \simeq 1.7-2.3$, similar to that found for other rotating MHD disk wind candidates such as HH212 ($\lesssim 5$; \citealt{Tabone2017}), HH30 ($\simeq 1.6$; \citealt{Louvet2018}), and DG Tau B ($\simeq 1.6$; \citealt{deValon2022}). We note that $\lambda_{\phi}$ is different from $\lambda_{\rm BP}$ (\citealt{Blandford1982}; \citealt{Tabone2020}). The latter measures the total specific angular momentum carried away by the wind which includes the magnetic torsion. While the former only measures the specific angular momentum for the matter rotation and cannot be higher than $\lambda_{\rm BP}$ \citep{Ferreira2006}. We take $\lambda_{\phi} \simeq 1.7$, consistent with the non-perturbed, inner data points in Figs. \ref{fig:diagnostic} and \ref{fig:diagnostic_03}, and thus a lower limit on $\lambda_{\rm BP}$ of 1.7. This low value of $\lambda_{\rm BP}$ is typically predicted for non-ideal MHD disk wind models with a relatively weak magnetization of $\beta \simeq 10^3-10^4$ in the disk midplane, where $\beta$ is the thermal-to-magnetic pressure ratio \citep{Lesur2021}. The corresponding disk surface density at 10~au is $\Sigma \simeq$ 100--600 g\, cm$^{-2}$, where we have used Eq.~(17) of \citealt{Lesur2021} with $M_\star = 0.4 \pm 0.1 M_{\odot}$ and a wind-driven disk accretion rate $\dot{M}_{\rm acc} \simeq (2-3) \times 10^{-6}\,M_{\odot}$\,yr$^{-1}$ (see Sect. 3.2). Combined with the measured disk radius of 23~au, it translates into a disk mass $M_d \simeq \pi R_d^2 \Sigma$ = 0.019-0.1 $M_{\odot}$, fully consistent with our independent estimate from dust continuum (see Sect. 3.1).

\subsection{Wind mass-flux}


Here we compare the wind ejection rate with the disk accretion rate to answer whether the wind can effectively drive accretion. The wind mass-loss rate was calculated by
 
\begin{equation}
     \dot{M}_{\rm DW} = 1.4 m_{\rm p} (N_{\rm HCN} dr) v_{z}/X_{\rm HCN}  
     \label{eq:mass-flux}, 
\end{equation}
 
\noindent where $m_{\rm p}$ is the proton mass, $v_z$ is taken as 2.5\,km\,s$^{-1}$ at a $z_{\rm proj}$ of 60\,au (see Fig. \ref{fig:all_v_r_z}), and the factor 1.4 is added to take into account for helium mass. The column density of HCN, $N_{\rm HCN}$, was calculated from that of H$^{13}$CN using the CASSIS spectral analysis tool (\citealt{Vastel2015}) by assuming a $^{12}$C/$^{13}$C ratio of 70 (\citealt{Milam2005}). The H$^{13}$CN spectrum was extracted in a rectangular aperture perpendicular to the jet axis with a length $dr=2\arcsec$ that encompasses all the emission in the lateral direction to the jet axis, and a width of ${\sim}$one beam ($0.1\arcsec$). It was centered at 60\,au from the continuum peak in the SE side. We note that because the H$^{13}$CN line profile is asymmetric, we fit two Gaussian profiles to find the column density using CASSIS. In this process, an excitation temperature of $100$\,K \citep{Belloche2020} was assumed, leading to an aperture-averaged column density of $N_{\rm{H^{13}CN}} = 7.2^{+0.8}_{-0.5}\times10^{13}$\,cm$^{-2}$, and  aperture-averaged line optical depth in each component ${\lesssim} 0.1$. At $z_{\rm proj}= 60$\,au, the flow width of ${\sim}0.4\arcsec$\, fills 1/5 of the aperture length ($2\arcsec$), hence the true line optical depth is ${\lesssim} 0.5$, and the optically thin assumption remains valid. The abundance of HCN with respect to the total hydrogen number density $n_{\rm H}$, $X_{\rm HCN}$, is assumed as $10^{-7}$ based on the measured column density ratios of HCN/CO in outflows from observations of \cite{Tafalla2010} and an assumed abundance of $10^{-4}$ for CO (e.g., \citealt{Pineda2008}). We note that $X_{\rm HCN}$ is relatively uncertain in MHD wind models and could have variations of ${\sim}1$ order of magnitude (e.g., \citealt{Gressel2020}), but $10^{-7}$ seem to be a good nominal value in observed protostellar outflows with smaller variation from source to source (a factor of ${\sim}3$; \citealt{Tafalla2010}). From our H$^{13}$CN observations, we estimate a mass-loss rate in the redshifted wind of $\dot{M}_{\rm DW} {\sim} 4 \times 10^{-6}$\,M$_{\odot}$\,yr$^{-1}$. Doubling to account for the blue lobe, the total mass-flux ($\dot{M}_{\rm DW, total}$) is ${\sim}2.5-4$ times larger than the accretion rate found in Sect. \ref{sec:envelope}.


Could this wind actually drive disk accretion at the estimated rate of $\dot{M}_{\rm acc} \simeq (2-3) \times 10^{-6}$\,M$_{\odot}$\,yr$^{-1}$? Assuming that all the angular momentum required to drive disk accretion is removed by the wind, \cite{Tabone2020} showed that the ratio of wind mass loss rate to disk accretion rate has to be (also see \citealt{Pascucci2023}) 
\begin{equation}
    f_{\rm M} \equiv \dot{M}_{\rm DW, total}/\dot{M}_{\rm acc}(r_{\rm in}) = \left(\frac{r_{\rm out}}{r_{\rm in}}\right)^{1/(2\lambda_{\rm BP} - 2)} - 1. 
    \label{eq:fM}
\end{equation}
\noindent where $r_{\rm in}$ and $r_{\rm out}$ are the inner and outer launch radii of the wind. In Eq. \eqref{eq:fM}, the lower the $\lambda_{\rm BP}$ the larger the $f_{\rm M}$ which means more mass is required to be launched in the wind to drive disk accretion. Similarly, for a larger wind launching region, more angular momentum needs to be extracted to advect gas from $r_{\rm out}$ to $r_{\rm in}$, increasing the mass-loss rate. The onion-like velocity structure seen in H$^{13}$CN suggests that the wind launch radius extends down to a much smaller radius than 10-20 au. The inner launch radius of the flow traced by H$^{13}$CN can be roughly estimated from the maximum projected velocity $\sim 12~$km~s$^{-1}$. Assuming that the inner streamline has $\lambda_{\phi} \simeq 1.7$, consistent with our non-perturbed, inner data points in Figs. \ref{fig:diagnostic} and \ref{fig:diagnostic_03}, and that the rotation velocity is similar to the axial velocity, as found for the outer streamline, we find from the Anderson relation that $r_{in} \simeq 1~$au. Using the derived values $r_{\rm in}=1$\,au, $r_{\rm out} = 10$\,au and a lower limit on $\lambda_{\rm BP}$ of 1.7 similar to $\lambda_{\phi}$, we find $f_{\rm M} {\sim} 4$, in agreement with the observationally derived mass-flux. We conclude that the total angular momentum carried away by the wind could be sufficient to drive disk accretion from the bulk part of the disk.


\section{Concluding remarks}
\label{sec:conclusions}

In this paper we presented high angular resolution (${\lesssim}0.1\arcsec$; ${\sim} 30$\,au) and high sensitivity ALMA observations of the L1448 protostellar system. Various components of the system (i.e., disk, inner envelope, and rotating outflow) are resolved for the first time. We attribute the rotating emission from CH$_3$OH and H$^{13}$CN at the base of the outflow to an MHD disk wind and suggest these two molecules as new tracers of an MHD disk wind for future observations. These two molecules are complimentary to the previously identified disk wind tracers such as SO, because these two are less contaminated by jet bowshocks and SO is more affected by chemistry and optical depth. Moreover, it is debated whether methanol, or more generally, complex organic molecules trace the inner warm envelope as sublimated from the ices or their emission structure is affected by shocks, disk winds, and presence of a disk. Here we find that methanol traces a rotating MHD disk wind candidate. Overall, our analysis opens new avenues to study the chemistry of disk-winds. Notably, understanding the formation, survival, and excitation of molecules such as SO, HCN, or methanol in those winds will be crucial to constrain the physics of disk-winds and access the composition of disks at the launching points of these winds.

We estimated the outer wind launch radius to be about ${\sim}10-20$\,au. The onion like structure of the flow indicates that the launching zone of the wind is extended down to ${\sim}1$ au. The magnetic lever arm parameter is relatively low ${\sim}1.7$, in line with MHD disk wind models from weakly magnetized disks ($\beta \simeq 10^3-10^4$). The mass-flux appears sufficient to drive disk accretion at the current rate. Detailed comparisons with MHD disk wind model predictions, and with alternative envelope entrainment models (e.g., \citealt{Rabenanahary2022}), are now required to obtain more robust constraints and definitely test the MHD interpretation. 





\begin{acknowledgements}
    We thank the referee for their constructive comments. We thank N. T. Kurtovic for the helpful discussions. Astrochemistry in Leiden is supported by EU A-ERC grant 101019751 MOLDISK, NOVA, and by the NWO grant 618.000.001. Support by the Danish National Research Foundation through the Center of Excellence “InterCat” (Grant agreement no.: DNRF150) is also acknowledged. B.T. and S.C. acknowledge support from the Programme National ``Physique et Chimie du Milieu Interstellaire'' (PCMI) of CNRS/INSU with INC/INP and co-funded by CNES. L.P. and C.C. acknowledge the project PRIN MUR 2022 FOSSILS (Chemical origins: linking the fossil composition of the Solar System with the chemistry of protoplanetary disks, Prot. 2022JC2Y93) and the INAF Mini-Grant 2022 “Chemical Origins” (PI: L. Podio). {\L}.T. acknowledges support from the ESO Fellowship Program. This paper makes use of the following ALMA data: ADS/JAO.ALMA\#2021.1.01578.S. ALMA is a partnership of ESO (representing its member states), NSF (USA) and NINS (Japan), together with NRC (Canada), MOST and ASIAA (Taiwan), and KASI (Republic of Korea), in cooperation with the Republic of Chile. The Joint ALMA Observatory is operated by ESO, AUI/NRAO and NAOJ. The National Radio Astronomy Observatory is a facility of the National Science Foundation operated under cooperative agreement by Associated Universities, Inc. 
\end{acknowledgements}

%
%

\bibliographystyle{aa}
\bibliography{wind}

\begin{appendix}




\section{A simple model for the outflow shape and kinematics}
\label{sec:toy_model}

\begin{figure}
  \resizebox{\columnwidth}{!}{\includegraphics{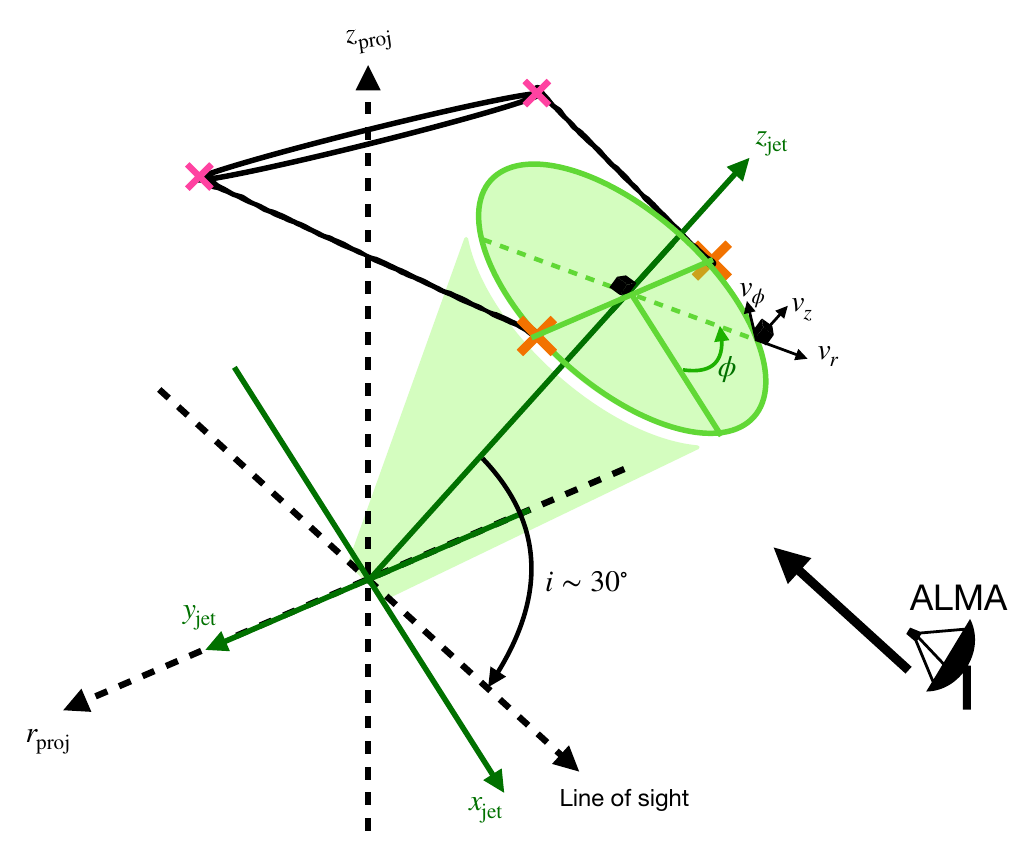}}
  \caption{Sketch of the outflow orientation relative to the plane of the sky. The black dashed lines (sky plane) are perpendicular to each other. The green solid lines (jet plane) are also perpendicular to each other. The orange crosses on the ring around the outflow walls correspond to the pink crosses on the solid circle which is the projection of the green ring onto the sky plane. Orange crosses are located at $\phi = \pm90^\circ$.}
  \label{fig:geo_wind}
\end{figure}

Here we explain a method to retrieve the velocities $v_\phi$, $v_r$, $v_{\rm z}$, and the shape of the outermost emitting layer of the outflow. This emitting layer may not correspond to a single streamline, due to chemical and excitation gradients. This method is applied to the redshifted lobe in Sect. \ref{sec:wind}. We show a sketch of the outflow geometry and define quantities used in the following in Fig. \ref{fig:geo_wind}. We assume that the outflow is axisymmetric. The outermost layer of the outflow can be divided into rings at different heights $z$ from the protostar (see one of these rings in light green in Fig. \ref{fig:geo_wind}). The velocity of this ring in the cylindrical coordinate system defined by the outflow axis is described by the rotation velocity $v_\phi$, the radial velocity $v_r$, and the axial velocity $v_{\rm z}$. We define the inclination angle $i$ as the angle between the blue lobe and the line of sight with $i=0$ for face-on and $i=\pi/2$ for edge-on configuration. In this convention, $v_{z}$ is positive in the blueshifted lobe and negative in the redshifted lobe, and the blueshifted lobe is mostly projected in the $z_{\rm proj}>0$ region of the sky. Based on the geometry presented in Fig. \ref{fig:geo_wind} a gas parcel located at a position ($r$, $z$, $\phi$) along the ring is projected on the plane of the sky at 
\begin{equation}
    r_{\rm proj} = - r\sin{\phi},
    \label{eq:rp}
\end{equation}
\begin{equation}
    z_{\rm proj} = z\sin{i} - r\cos{i}\cos{\phi},
    \label{eq:zp}
\end{equation}
\noindent and has a projection velocity (also often called "radial velocity") 
\begin{equation}
    v_{\rm proj} = - v_{z}\cos{i} -v_{r}  \sin{i}\cos{\phi} +v_{\phi} \sin{i}\sin{\phi},
    \label{eq:vp}
\end{equation}  
\noindent where subscript proj indicates `projected' and where positive velocities stand for redshifted material.

We aim to retrieve the outer radius of the flow $r$ and the three velocity components as a function of $z$ from the transverse position-velocity (PV) diagrams. The challenge comes from the fact that even at high angular resolution, a transverse PV diagram contains the emission from a collection of gas parcels located at different positions ($r$, $z$, $\phi$).
However, at given $z_{\rm proj}$ the maximal projected radii beyond which no emission is detected in the PV diagram, denoted as $r_{\rm proj,1}$ ($>0$) and $r_{\rm proj,2}$ ($<0$), correspond to gas parcels located at $\phi = - 90^\circ$ and $+90^\circ$, respectively, $r= r_{\rm proj,1} \simeq - r_{\rm proj,2}$ and $z=z_{\rm proj}/\sin{i}$ (see the two orange crosses in Fig. \ref{fig:geo_wind}). The shape of the outer layer of the outflow $r(z)$ can therefore be directly inferred from the measurement of $r_{\rm proj,1}$ and $r_{\rm proj,2}$ at different distances $z_{\rm proj}$ as
\begin{align}
\begin{split}
    r & = \frac{|r_{\rm proj,1}|+|r_{\rm proj,2}|}{2} \\
    z & = \frac{z_{\rm proj}}{\sin{i}}.
    \label{eq:shape_estimates}
\end{split}
\end{align}

Regarding the kinematics of the outflow, we note that the gas parcels emitting at $r_{\rm proj,1}$ and $r_{\rm proj,2}$ have projected velocities of $v_{\rm proj,1} = - v_z \cos{i} - v_{\phi} \sin{i}$ and $v_{\rm proj,2} = - v_z \cos{i} + v_{\phi} \sin{i}$. Therefore the velocities found in the transverse PV diagram at $r_{\rm proj,1}$ and $r_{\rm proj,2}$ give the rotation and axial velocity at a distance $z = z_{\rm proj}/\sin{i}$ of
\begin{equation}
\begin{split}
    v_{\phi} & = \frac{v_{\rm proj,2} - v_{\rm proj,1} }{2 \sin{i}} \\
    v_z      & = - \frac{v_{\rm proj,1} + v_{\rm proj,2} }{2 \cos{i}}
\label{eq:velocity_estimates}
\end{split}
\end{equation}



\begin{table}
\renewcommand{\arraystretch}{1.3}
    \caption{P-V measurements}
    \label{tab:r_z_meas}
    \resizebox{\columnwidth}{!}{\begin{tabular}{@{\extracolsep{1mm}}*{5}{l}}
          \toprule
          \toprule    
        Observable & $z_{\rm proj}\,[\rm au]$&&&\\
        &30&  60 & 90 & 120 \\ 
        \midrule     

$r_{\rm proj,1}$\,[$\arcsec$]& 0.10(0.05)& 0.12(0.05)& 0.20(0.05)& 0.25(0.05)\\
&0.08(0.05)& -&-&-\\
$v_{\rm proj,1}$\,[km\,s$^{-1}$]& -0.9(1)&0.8(1)&2.0(1)& 2.3(1)\\
&-1.1(0.5)&-&-&-\\
$r_{\rm proj,2}$\,[$\arcsec$]& -0.08(0.05)& -0.10(0.05)& -0.19(0.05)& -0.25(0.05)\\
&-0.1(0.05)&-&-&-\\
$v_{\rm proj,2}$\,[km\,s$^{-1}$]& 3.1(1)& 3.7(1)& 5.1(1)& 5.5(1)\\
&2.5(0.5)& -&-&-\\

\bottomrule
        \end{tabular}}
        \tablefoot{Values in parenthesis give the uncertainties. The first row for each observable is measured from H$^{13}$CN P-V cuts and the second row from CH$_3$OH P-V cuts.}
\end{table}

\begin{figure*}
    \centering
    \includegraphics[width=15cm]{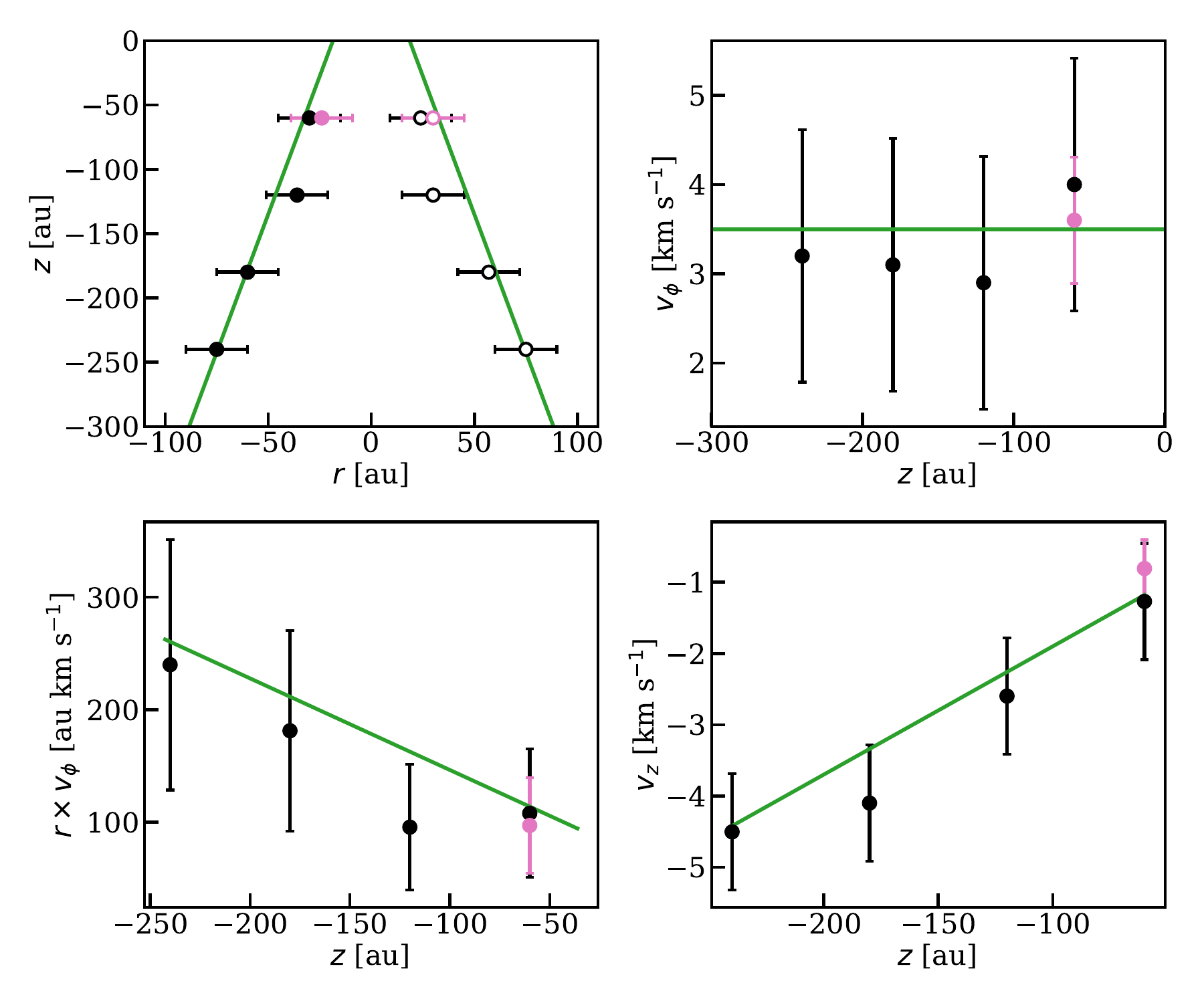}
    \caption{Retrieved shape and kinematics of the wind outermost emitting layer in the red lobe. Radius $r$ (top left), velocities $v_\phi$ (top right) and $v_{z}$ (bottom right), and specific angular momentum $r \times v_\phi$ (bottom left) are shown as a function of deprojected altitude $z$. Data points with error bars represent values for CH$_3$OH (pink) and H$^{13}$CN (black), derived from the PV-measurements in Table~\ref{tab:r_z_meas} and deprojected by an inclination $i = 30\degr$. The values of $r \times v_{\phi}$ are computed using the average flow radius between the two edges. Green lines show our simple conical model best fitting the PV diagrams in Fig. \ref{fig:PVs}, assuming a flow along the cone walls.} 
    \label{fig:all_v_r_z}
\end{figure*}


We report in  Table \ref{tab:r_z_meas} the values of $r_{\rm proj,1/2}$ and $v_{\rm proj,1/2}$ measured for H$^{13}$CN and CH$_3$OH at various $z_{\rm proj}$. 
The shape of the outflow, $v_\phi$, and $v_{z}$ were then calculated from Eqs. (\ref{eq:shape_estimates}) and (\ref{eq:velocity_estimates}).
Finally, we find $v_r$ by assuming that the poloidal velocity is tangential to the flow shape. Figure \ref{fig:all_v_r_z} presents the measured values of $v_z$, $v_\phi$, and $rv_\phi$ as a function of $z$. 

In order to check the method, we compute the location of the emission in each PV diagram (see black lines in Fig. \ref{fig:PVs}), accounting for the contribution of all the gas parcels and not only those at $\phi = \pm 90^{\circ}$. In order to do so, one needs to prescribe $r$, $v_{\phi}$, $v_{z}$, and $v_{r}$ as function of $z$. These relations are obtained by fitting the shape of the outflow by a cone and the $v_{z}$ by a linear function of $z$. We assume $v_{\phi}$ to be constant and equal to +3.5\,km\,s$^{-1}$.
During this process, we played with the various parameters to find the best-fit ring-like structure for the P-V diagrams. The best fit to the PV diagrams in Fig. \ref{fig:PVs} results in a relation between $r$ and $z$, and $v_z$ and $z$ of $(z/1~\rm au)= \pm 4.3 (r/1~\rm au) + 80$ and $(v_{z}/1\,\rm km s^{-1}) = 0.018(z/1~\rm au) - 0.1$, respectively (green curves in Fig. \ref{fig:all_v_r_z}).





\section{Additional plots}
\label{app:add_plots}
\begin{figure*}
    \centering
    \includegraphics[width=19cm]{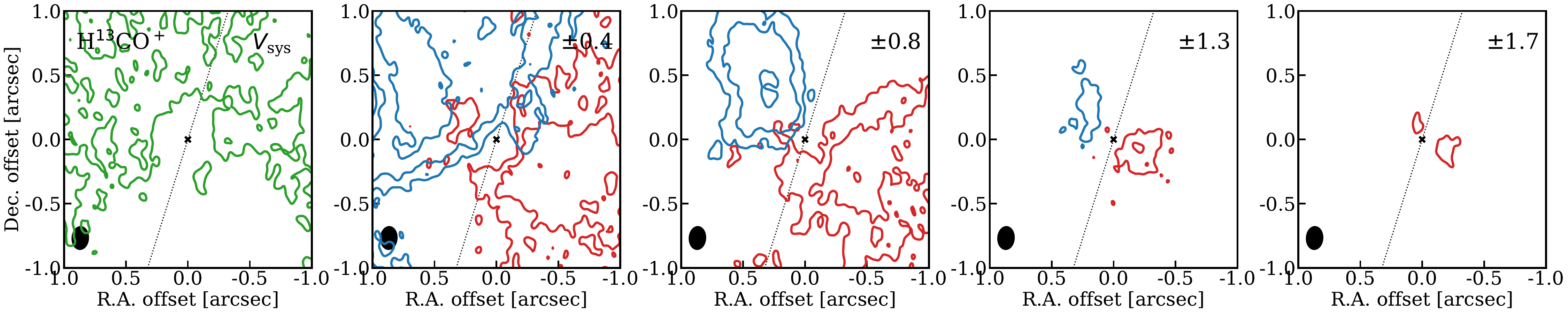}
    \caption{Channel maps of H$^{13}$CO$^+$ 4-3. To create this image the cube was re-binned in the spectral axis by a factor of two to increase S/N. The contours are at levels [6, 9, 14]$\sigma$ with $\sigma$ of this re-binned cube being equal to 1.8\,mJy\,beam$^{-1}$. The red and blue contours correspond to positive and negative velocities respectively, expressed in km\,s$^{-1}$ with respect to the adopted systemic velocity $V_{\rm sys}$ (= 5.3 km\,s$^{-1}$ in the LSR frame). The peak of the continuum is indicated by a cross and the jet axis by a dotted line.} 
    \label{fig:HCO}
\end{figure*}

\begin{figure*}
    \centering
    \includegraphics[width=19cm]{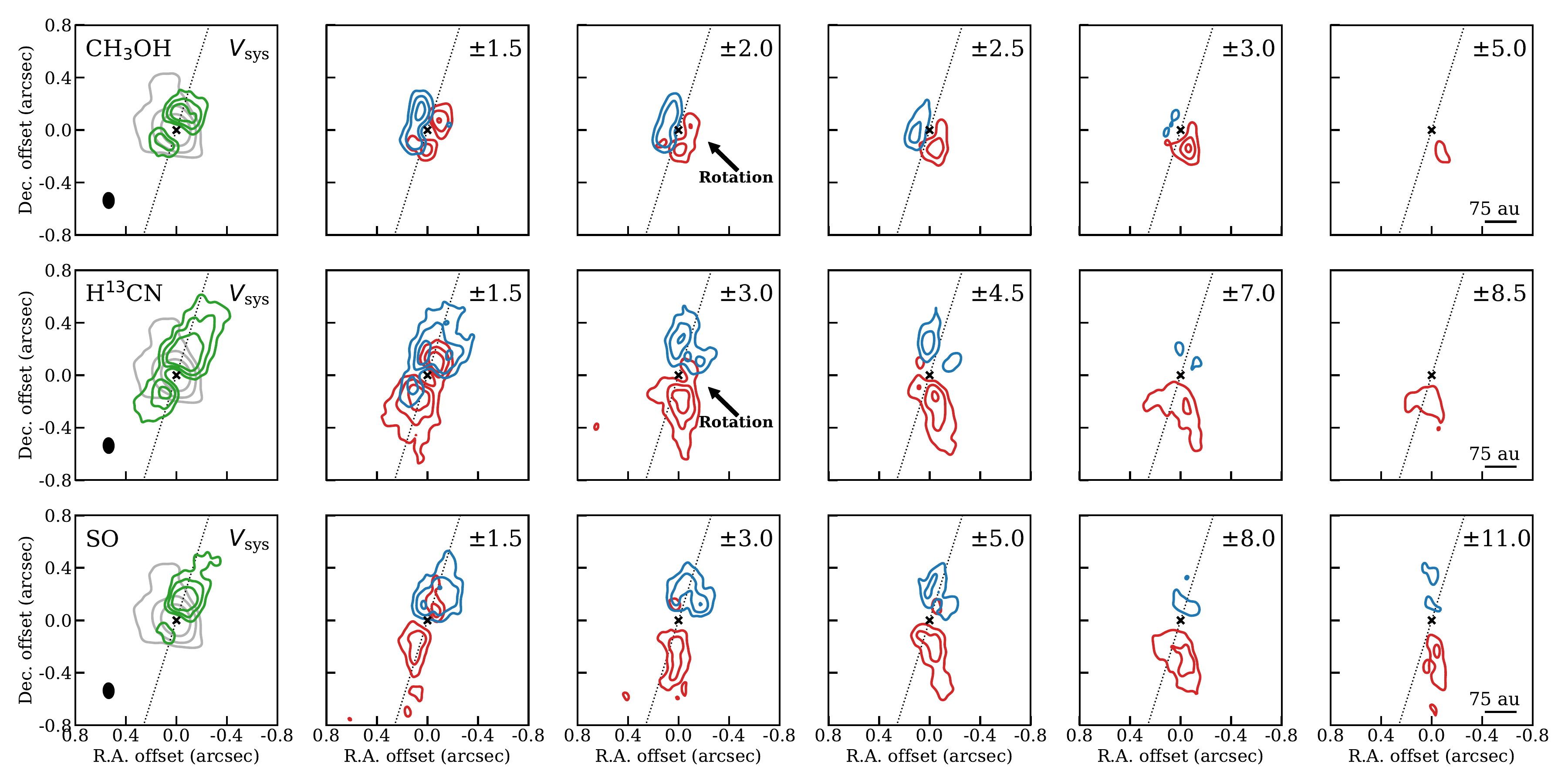}
    \caption{Channel maps of CH$_3$OH (2$_{2,1,0}-3_{1,2,0}$), H$^{13}$CN (4-3), and SO (8$_9-7_8$) in the inner ${\sim} 1.5\arcsec$ region. The continuum contours are shown in gray and they are set at [10, 30, 100]$\sigma_{\rm cont}$ with $\sigma_{\rm cont} = 0.25$\,mJy\,beam$^{-1}$. The peak of the continuum is indicated by a cross. The red and blue contours correspond to positive and negative velocities respectively, expressed in km\,s$^{-1}$ with respect to the adopted systemic velocity $V_{\rm sys}$ (= 5.3 km\,s$^{-1}$ in the LSR frame). The contour levels are set at [7, 13, 20]$\sigma$ for CH$_3$OH and H$^{13}$CN, and [6, 10, 15]$\sigma$ for SO with $\sigma$ given in Table \ref{tab:obs}. The molecules shown here show rotation signatures at $|V| < 3$\,km\,s$^{-1}$ in the inner regions or in the outflow emission.} 
    \label{fig:chans_all}
\end{figure*}

\begin{figure}
  \resizebox{\columnwidth}{!}{\includegraphics{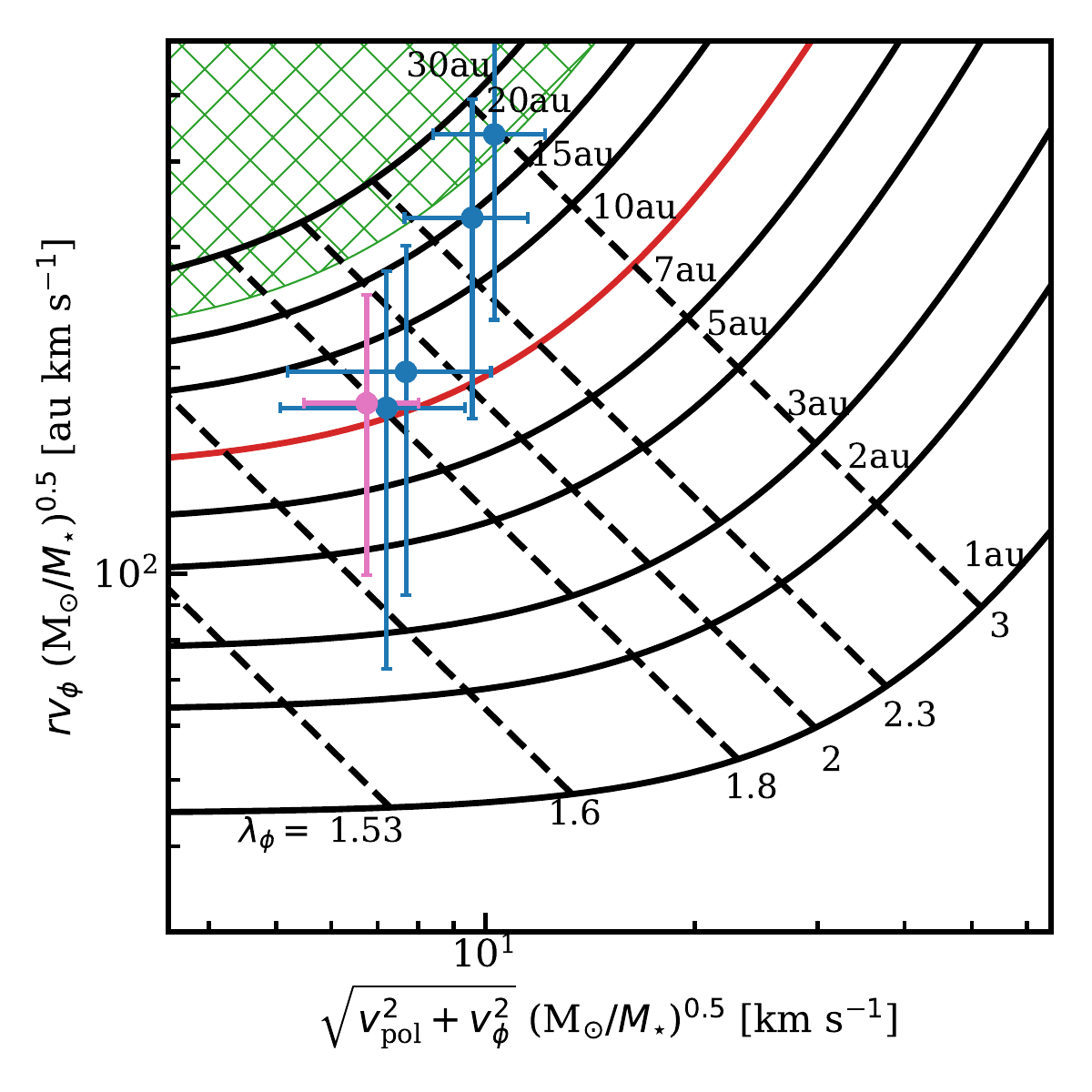}}
  \caption{Same as Fig. \ref{fig:diagnostic} but assuming a protostellar mass of $0.3$\,M$_{\odot}$.}
  \label{fig:diagnostic_03}
\end{figure}

Figure \ref{fig:HCO} presents the channel maps of H$^{13}$CO$^+$ in the L1448-mm system. Figure \ref{fig:chans_all} presents the channel maps of CH$_3$OH (2$_{2,1,0}-3_{1,2,0}$), H$^{13}$CN (4-3), and SO (8$_9-7_8$) to demonstrate the rotation in their outflow emission. Figure \ref{fig:diagnostic_03} is the same as Fig. \ref{fig:diagnostic} but for a protostar with a mass of 0.3\,M$_{\odot}$.


\end{appendix}

\end{document}